\documentclass[twocolumn,runningheads]{svjour2}
\smartqed  
\usepackage{graphicx}
\journalname{Astrophysics and Space Science}
\begin{document}

\title{Transient X-ray sources in the field of the Unidentified Gamma-Ray
Source TeV J2032+4130 in Cygnus
}
\titlerunning{Transient X-ray sources in the field of TeV J2032+4130}        

\author {R. Mukherjee \and 
E. V. Gotthelf \and 
J. P. Halpern 
}


\institute{R. Mukherjee \at 
              Department of Physics \& Astronomy,
              Barnard College, Columbia University, 
              New York, NY 10027.       \\        
\email{muk@astro.columbia.edu}
           \and
           E. V. Gotthelf \& J. P. Halpern \at 
Columbia Astrophysics Laboratory, 
Columbia University, 
New York, NY 10027
}

\date{Received: date / Accepted: date}

\maketitle

\begin{abstract}
We present an analysis of {\it Chandra} ACIS observations of the field of TeV
J2032+4130, the first unidentified TeV source, detected serendipitously by
HEGRA. This deep (48.7 ksec) observation of the field follows up 
on an earlier 5 ksec {\it Chandra} director's discretionary observation. Of the
numerous point-like X-ray sources in the field, the brightest are shown to be 
a mixture of early and late-type stars. We find that several of the X-ray
sources are transients, exhibiting rapid increases in count rates by factors
3--10, and similar in nature to the one, hard absorbed transient
source located in the earlier {\it Chandra} observation of
the field. 
None of these transient sources are likely to correspond to the TeV source. 
Instead, we identify a region of diffuse
X-ray emission within the error circle of the TeV source and consider its
plausible association. 

\end{abstract}

\section{Introduction}
\label{intro}

TeV J2032+4130 was the first unidentified gamma-ray source detected at
TeV energies. The source was discovered serendipitously in the direction
of the Cygnus OB2 stellar association region by the HEGRA CT-System at
La Palma \cite{Ref1,Ref2} in
observations originally devoted to Cygnus X--3. The HEGRA
observations, carried out between 1999 and 2002, found TeV J2032+4130
to be a steady gamma-ray source, with the integrated flux measured
above 1~TeV at $\approx 5\%$ of that of the Crab Nebula. The best fit
HEGRA position for the source is $20^{\rm h}31^{\rm
  m}57^{\rm s}\pm 6.\!^{\rm s}2_{\rm stat}\pm1.\!^{\rm s}0_{\rm sys},$ 
$+41^{\circ}29^{\prime}56.8^{\prime\prime} \pm 1.\!^{\prime}1_{\rm
  stat} \pm 1.\!^{\prime}0_{\rm sys}$ (J2000) \cite{Ref3}. The source was reported to be extended, with a Gaussian
$1\sigma$ radius of $\sim 6.\!^{\prime}2\ (\pm 1.\!^{\prime}2_{\rm
  stat} \pm 0.\!^{\prime}9_{\rm sys})$. TeV J2032+4130 was also
detected in the Whipple archival data taken in 1989 and 1990 \cite{Ref4}, 
with some indication that the source may be variable on the
time scale of several years. It is interesting to
note that the error circle of TeV J2032+4130 overlaps the edge of the
95\% confidence error ellipse of an EGRET source, 3EG
J2033+4118. However, it is not clear if they are associated. 

The field of TeV J2032+4130 was initially observed by {\it Chandra}
during a short 5 ksec exposure \cite{Ref5}. 
In an attempt to understand the possible origins of the source, we
performed a multiwavelength study of the region, carrying out
optical identifications and spectroscopic classifications of the
bright X-ray sources in the {\it Chandra} ACIS image and (archival)
{\sl ROSAT} PSPC data \cite{Ref6}. The X-ray sources
detected were found to be a mix of early- and late- type stars, and
there was no compelling counterpart to the gamma-ray
source. However, in our study of the {\it Chandra} ACIS field, we did
find an unusual new, hard absorbed source that was 
both transient and rapidly variable. We reported on the detection of this
source ({\it Chandra} Source 2 in \cite{Ref6}) as the
brightest source in the {\it Chandra} field, located at 
$20^{\rm h}31^{\rm  m}43.755^{\rm s},$ $+41^{\circ}35^{\prime}55.17^{\prime\prime}$ (J2000), at a distance of $7^\prime$ from the centroid of the TeV
emission. We detected a coincident reddened optical counterpart with
the MDM Observatory 2.4m telescope, but without any emission or absorption
features in its spectrum. Although the transient source was the
brightest of the {\it Chandra} sources, it was noticeably absent from
earlier {\sl ROSAT} or {\it Einstein} images. 
At the time of our initial study of this field, we
considered the possibility of this transient X-ray source being a
candidate for a ``proton'' blazar \cite{Ref7}, a radio-weak gamma-ray
source that could be associated with TeV J2032+4130 \cite{Ref6}. 
However, without knowing the exact nature of this source,
we were unable to consider it a compelling counterpart. 

Since the original study, a deep 50 ksec {\it Chandra}
observation of the TeV J2032+4130 field has been acquired. An analysis
of this data recently summarized in \cite{Ref8} found $\approx
240$ point-like X-ray
sources in the {\it Chandra} field, but no obvious diffuse X-ray
counterpart to TeV J2032+4130. 
We have reanalyzed this new {\it Chandra} exposure and
find that at least seven of the brightest 
X-ray point sources are either flare stars or
transients. The brightest {\it Chandra} source, ``Source 2'' from our
earlier study is no longer detected. We are now convinced that {\it
  Chandra} Source 2 reported in \cite{Ref6} is a flare
star, not associated with TeV J2032+4130. In this paper we (a) summarize
the properties of the several transient sources discovered in the {\it Chandra} field of TeV
J2032+4130, (b) identify a diffuse candidate X-ray counterpart, and (c) review our conclusions about the possible nature of
the gamma-ray source.  

\section{X-ray Observations}
\label{sec:1}

On 2004 July 12, {\it Chandra} acquired a 48.7 ks observation 
of the field of TeV J2032+4130 with the front-illuminated, imaging CCD array of
the Advanced CCD Imaging Spectrometer 
(ACIS-I). ACIS is sensitive to photons in the energy range 0.2--10 keV with a 
spectral resolution of $\Delta E/E\sim 0.1$ at 1 keV. Data reduction and
analysis were performed using the standard analysis software packages, CIAO, 
FTOOLS, and XSPEC. 
Figure~1 shows the {\it Chandra} image of the region, with the
position of TeV J2032+4130 marked. There are numerous 
pointlike X-ray sources near the centroid of the TeV
source: Butt et al. (2006) find 240 point-like
X-ray sources in a recent study of the field \cite{Ref8}. We have marked the positions of the
brightest point sources, those having at least 100 photons and a signal-to-noise
ratio $5\sigma$ or greater, in Fig. 1. The positions, count rates and hardness
ratios of these sources are given in Table 1. 
A comparison of this field with the earlier 5 ksec {\it Chandra} 
exposure \cite{Ref6} shows that several of the point sources from the earlier {\sl
  Chandra} observation are detected in this deeper exposure. 
However, it is notable that the brightest X-ray source from the  earlier
observation is absent from the image shown in Fig. 1. The
position of this transient source discovered by \cite{Ref6} is marked in the figure with a triangle. 

\begin{figure}
  \begin{center}
    \includegraphics[height=25pc]{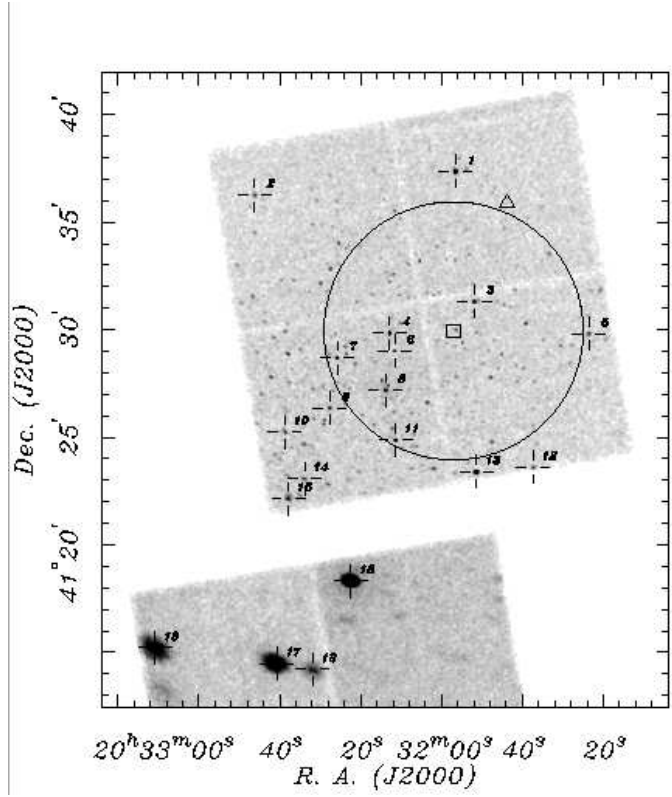}
  \end{center}
  \caption{{\it Chandra} ACIS-I image of the field of TeV J2032+4130.
The positions of the marked sources are given 
in Table 1. The small square marks the centroid of TeV J2032+4130, and the circle
is the estimated Gaussian $1\sigma$ extent of the TeV emission \cite{Ref3}.
The triangle marks the brightest {\it Chandra} source in an earlier 5
ksec observation of the region \cite{Ref6}, noticeably
absent from this image. }
\label{fig1}
\end{figure}

We find that several of the sources in the {\it Chandra} field have ordinary
stellar counterparts. Many of the stars in the Cyg OB2 association are among
the strongest stellar X-ray sources in the Galaxy. 
Likely optical identifications of the {\it Chandra}
sources are given in Table 1. The magnitudes listed in the table are from the
USNO-A2.0 and USNO-B1.0, where available, or from the MT91 \cite{Ref9} 
compilation of stars in Cyg OB2, or from the optical images obtained by us
during our earlier study of this field \cite{Ref6}.  Two of the 
sources have no optical counterparts. Both were detected in 
the earlier {\it Chandra} observation, and have no optical counterpart to a 
limiting magnitude greater than 23 \cite{Ref6}. As in our earlier analysis, we find these to 
be the hardest sources in the image. They are likely to be active galaxies, highly absorbed 
by the Galactic ISM, and are unlikely to be nearby, old neutron stars.  

\begin{table*}[t]
\caption{{\it Chandra} Sources in the Field of TeV J2032+4130: Likely
Counterparts}
\begin{center}
\label{tab:1}
\begin{tabular}{lclclllcc}
\hline\noalign{\smallskip}
ID &  X-ray Position$^a$ & Cts$^b$& 
HR$^c$ & Optical Position$^a$ & Name & Sp. & $B$ & $R$ \\
 & R.A.\hskip 4em Decl. &  &  & R.A.\hskip 4em Decl. & & Type & mag & mag\\[3pt]
\tableheadseprule\noalign{\smallskip}
 1 & 20 31 56.50 +41 37 22.00 &  808   &  0.56  & . . .   . . .             & . . .      &     &        &$>23.2$\\
 2 & 20 32 46.23 +41 36 16.03 &  445   & $-$0.22& 20 32 46.240 +41 36 16.0  & MT91 321   &     & 11.23  &10.28 \\
 3 & 20 31 51.87 +41 31 18.91 &  206   &  0.76  & . . .   . . .             & . . .      &     &        &$>23.7$\\
 4 & 20 32 12.78 +41 29 50.94 &  843   &  0.15  & 20 32 12.763 +41 29 51.24 & . . .      &     &        &19.2\\
 5 & 20 31 23.58 +41 29 49.29 &  274   & $-$0.21& 20 31 23.573 +41 29 49.45 & . . .      &     &        &15.1\\
 6 & 20 32 11.60 +41 29 01.41 &  113   & $-$0.05& 20 32 11.600 +41 29 01.48 & . . .      &     &        &15.5\\
 7 & 20 32 25.78 +41 28 42.28 &  160   & $-$0.18& 20 32 25.731 +41 28 42.89 & . . .      &     &        & 17.3\\
 8 & 20 32 13.84 +41 27 11.66 &  290   & $-$0.75& 20 32 13.836 +41 27 12.33 & Cyg OB2 4  &O7 III((f))&11.42&10.2\\
 9 & 20 32 27.63 +41 26 21.76 &  179   & $-$0.60& 20 32 27.663 +41 26 22.44 & MT91 258   &     &        &10.4 \\
10 & 20 32 38.72 +41 25 14.75 &  440   & $-$0.31& 20 32 38.580 +41 25 13.6  & MT91 299   &O7.5V& 12.03  &    \\
11 & 20 32 11.32 +41 24 52.02 &  301   & $-$0.08& 20 32 11.303 +41 24 52.69 & . . .      &     &        & 17.0\\
12 & 20 31 37.32 +41 23 37.19 &  232   & $-$0.51& 20 31 37.267 +41 23 36.01 & MT91 115   & G6 V& 13.90  &13.1\\ 
13 & 20 31 51.30 +41 23 23.44 &  722   & $-$0.78& 20 31 51.319 +41 23 23.79 & MT91 152   & G3 V& 13.40  &13.1\\
14 & 20 32 33.84 +41 23 04.46 &  623   &  0.17  & 20 32 33.862 +41 23 04.27 & . . .      &     &        &16.7\\
15 & 20 32 37.85 +41 22 08.79 &  931   &  0.10  & 20 32 37.820 +41 22 08.98 & . . .      &     &        &18.5\\
16 & 20 32 22.42 +41 18 18.97 & 28127  & $-$0.55& 20 32 22.425 +41 18 18.96 & Cyg OB2 5  & O7e & 10.64  & 8.1\\
17 & 20 32 40.66 +41 14 28.96 & 16558  & $-$0.26& 20 32 40.959 +41 14 29.29 & Cyg OB2 12 & B5Iab:&14.41 & ...\\
18 & 20 32 31.85 +41 14 12.15 &  9440  & $-$0.13& 20 32 31.556 +41 14 08.48 & MT91 267   & ... & 15.06  &11.8\\ 
19 & 20 33 10.83 +41 15 12.56 & 14479  & $-$0.28& 20 33 10.736 +41 15 08.22 & Cyg OB2 9  & O5Iab:e & 12.61 & ...\\
\noalign{\smallskip}\hline
\end{tabular}
\end{center}

$(a)$ Units of right ascension are hours, minutes, and seconds.
Units of declination are degrees, arcminutes, and arcseconds. 
$(b)$ Total counts in a $12^{\prime\prime}$ radius aperture. 
The total included background is estimated as 1--3 counts. 
$(c)$ Hardness ratio (HR) is defined as: ${S(0.5-2 {\rm keV}) - S(2-10
{\rm keV})}\over{S(0.5-2 {\rm keV}) + S(2-10
{\rm keV})} $, where $S$ is the source counts in a given energy band. 
\end{table*}

\section{Transient X-ray Sources in the Field of TeV J2032+4130}
\label{sec:2}
Seven of the sources in Table 1 were not detected in the earlier 
{\sl ROSAT} \cite{Ref10} or {\it Chandra} \cite{Ref6} 
observations of this field. Thus, they may be described as transient sources. 
Figure 2 shows the lightcurve of the brightest of these sources (\# 4), constructed 
from the 48.7 ks {\it Chandra} observation. The aperture of the source and 
background regions are indicated in the figure. The background is seen to be
sufficiently stable and has little effect on the source light curves. The figure
shows that the count rate rose by more 
than a factor of 10 in the final 15 ks of the observation, after remaining 
faint for the first 35 ks.  We see a similar behavior in the case of the other
transient sources. We believe that these are flare stars, which commonly exhibit 
X-ray flaring activity (e.g. \cite{Ref11}). 

\begin{figure}[b]
  \begin{center}
    \includegraphics[height=20pc,angle=-90]{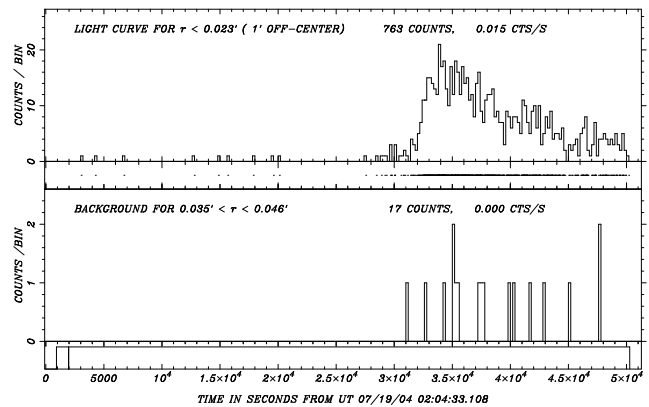}
  \end{center}
  \caption{Lightcurves (top) and local background (bottom) of {\it Chandra}
source 4 (Table 1). The aperture sizes for
the source extraction and background regions are indicated in the figure. A time
binning of 200 s was used. The dots below the light curve are the arrival times
of the individual photons. The background is demonstrated to be stable and has
negligible effect on the source light curve.}
\label{fig:2}
\end{figure}

In comparison, the only transient source in our earlier {\it Chandra} field was also
highly variable during the brief 5 ks observation. It remained faint for the first 
3.5 ks, but increased its count rate by ten-fold in the final 1.5 ks. This
source is not detected in the deep {\it Chandra} observation of the field. 

Flare stars are generally dim, red (class-M) dwarfs  
that are seen to exhibit unusually violent activity in optical and/or
X-ray bands, and sometimes in the radio and ultraviolet bands. 
Flare stars are not known to
be gamma-ray emitters. It is unlikely that any of the transient X-ray sources
are point source counterparts to TeV J2032+4130. 

One of the suggestions for the origin of an extended region of TeV emission, as
in the case of TeV J2032+4130, is 
inverse Compton scattering from a jet-driven termination shock from Cyg X-3 or
an as yet undetected microquasar \cite{Ref1}. In \cite{Ref6} we were motivated to further study the one transient source in the
field in order to investigate if it could be such a jet source. Based on the 
new {\it Chandra} observation and the detection of several similar transient
sources, it is clear that the one transient source in \cite{Ref6} 
is not a microquasar, or responsible for the TeV
emission in any way. 

\section{Diffuse X-ray Emission in the Field of TeV J2032+4130}
\label{sec:3}

We carried out an analysis to search for diffuse, extended X-ray emission in 
the TeV source region. Figure 3 shows an image of the diffuse emission only,
made by 
locating and cutting out the point sources, and smoothing the resulting image
with a Gaussian kernel of sigma $14^{\prime\prime}$. This image is exposure and
vignetting corrected in the broad energy band 
of 0.3 to 8 keV. The extended emission centered on $20^{\rm h}32^{\rm
m}13.4^{\rm s},\ +41^{\circ}27^{\prime}10.4^{\prime\prime}$ (J2000) is detected
at a significance of $6.1\sigma$, and has an extent of roughly $\sim 1.6^\prime$
diameter, with a few features extending further. It is possibly associated with
Cyg OB2 \#4. For the circular aperture of $1.6^\prime$ diameter, the
diffuse X-ray flux is $8\times 10^{-14}$ erg cm$^{-2}$ s$^{-1}$ in the 0.5--10
keV band, assuming a power-law spectral model with photon index of 1.5, typical
of non-thermal spectra and total Galactic $N_H = 1.5\times 10^{22}$ cm$^{-2}$. 
By comparing the exposure and vignetting corrected images of the region in
the soft and hard energy bands of 0.3--2.0 keV and 2.0--8.0 keV,
respectively, we find no significant softening of the spectrum in the high
energy band. The corresponding hardness ratio is $-$0.48. 

\begin{figure}[t]
  \begin{center}
    \includegraphics[height=17pc]{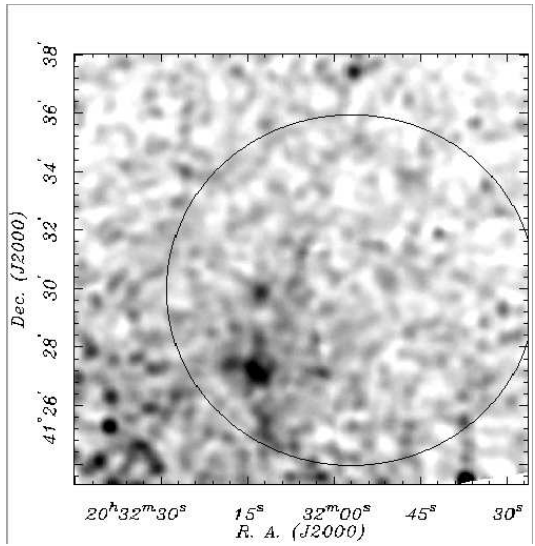}
  \end{center}
  \caption{{\it Chandra} ACIS-I image of the field of TeV J2032+4130 showing
diffuse emission only in the 0.3-8 keV band. The image has been exposure and
vignetting corrected. The circle
is the estimated Gaussian $1\sigma$ extent of the TeV emission \cite{Ref3}. }
\label{fig:3}
\end{figure}

\section{Discussion and Conclusions}
\label{sec:4}

Based on the deep {\it Chandra} observations of the TeV field, we summarize our
principal findings as follows. 

We find several new transient sources in the {\it Chandra} field of TeV
J2032+4130. These are similar in nature to the one transient source found in 
the
earlier 5 ks {\it Chandra} observation of the field. We are convinced that 
these transient sources are 
flare stars, unlikely to be associated with the TeV source.  

Mukherjee et al. (2003) \cite{Ref6} considered the candidacy of the one transient source detected in the
field of TeV J2032+4130 for either a ``proton blazar'' or a jet source
responsible for TeV emission via inverse Compton scattering. Based on the new
data, we are convinced that this is not the case. 

We find no convincing point source counterpart to TeV J2032+4130 in the X-ray
band. 

We find significant hard diffuse X-ray emission within the error circle of TeV
J2032+4130. If the source of the diffuse emission is embedded in the Cygnus OB2
association at $d=1740$ pc \cite{Ref9}, the corresponding luminosity is $\sim 3\times
10^{31}$ erg s$^{-1}$.

TeV J2032+4130 appears to be an extended source, unlikely to have a 
point source counterpart at other wavelengths.  It seems to be related to the
massive Cyg OB2 association and the massive stars in the region. 
Aharonian et al. (2002) \cite{Ref1} 
discuss two possible origins of the extended TeV emission from the source. The
emission could be hadronic in origin, arising from the acceleration of hadrons
in shocked OB star winds and interaction with local, dense gas cloud, and
subsequent $\pi^0$ decay. 
Or, the TeV emission could be inverse Compton scattering in a jet-driven
termination shock from Cyg X-3  or an as yet undetected microquasar. 

In a recent study, Butt et al. (2006) \cite{Ref8} find that a 
surface density plot of the point-like X-ray sources in the {\it Chandra} field
shows an excess consistent 
with the size and position of TeV J2032+4130. One proposal made by these authors
is that the TeV source is a composite of several point sources, and it is
possible that several of point X-ray sources are responsible for the TeV
emission. 

The fact that we detect hard X-ray emission within the error circle of TeV
J2032+4130 is quite interesting. Together with the TeV observations, it points
to the fact that high energy particles are being accelerated in the stellar
winds associated with the massive stars in the region. It is not obvious, however, that the
diffuse emission is related to the TeV source. We need deeper observations
of the region in order to derive an X-ray spectrum of the diffuse
emission. It would also be important to see if future observations (with better
angular resolution) of TeV
J2032+4130 with 
VERITAS or MAGIC indicate any spatial correlation between the gamma-ray and
X-ray emissions.  Further observations at TeV energies with ground-based
atmospheric Cherenkov telescopes as well as space-based experiments like 
GLAST are needed to help us resolve the nature of this source. 

\begin{acknowledgements}

This publication makes use of data obtained from HEASARC at Goddard Space Flight Center 
and the SIMBAD astronomical database.  
R. M. acknowledges support from NSF grant PHY-0244809. 

\end{acknowledgements}


\end{document}